\documentclass[a4paper]{article}

\usepackage{graphicx,amsmath,natbib,a4wide}

\newcommand{\ignore}[1]{}

\begin{document}

\title{A multifractal model for spatial variation in species richness}\date{}

\author{Henri Laurie \\
Mathematics and Applied Mathematics, University of Cape Town\\ 
\tt{henri.laurie@uct.ac.za}\\  \\ 
Edith Perrier \\
UR GEODES, IRD, Bondy \\ 
\tt{perrier@ird.bondy.fr}
 }

\maketitle
\enlargethispage{2cm}

Keywords: species-area relationship, multifractal, power-law, spatial scaling, spatial variability

\newpage

\begin{abstract} Models for species-area relationships up to now have focused on the mean richness as a function of area. We present MFp1p2, a self-similar multifractal. It explicitly models both trend and variation in richness as a function of area, and is a generalisation of the model of scaling of mean species richness due to \citet{hartetal99:334}. The construction is based on a cascade of bisections of a rectangle. The two parameters of the model are $p_1$, the proportion of species that occur in the richer half, and $p_2$, the proportion of species that occur in the poorer half. Equivalent parameterisations are $a = (p1 + p2)/2$ and $b = p1/p2$. These parameters are interpreted as follows: $a$ gives the scaling of mean density, $b$ gives the scaling of spatial variability. Several properties of MFp1p2 are derived, a generalisation is noted and some applications are suggested.

\end{abstract} 

\section*{INTRODUCTION}

Ecologists study organismal interaction on a wide range of scales. From the centimetres of a puddle to the megametres of the planet, the characteristic length of the system spans some nine orders of magnitude. The mere description of pattern over such a range of scales is a major challenge. It becomes formidable if in addition one requires that the description be simple.

But species-area relationships (SARs) achieve exactly that. Let $S$ denote the number of species found in a survey of a spatial domain with area $A$. Then one may hope that $S$ increases with $A$ in a regular way. Indeed, under the interpretation that $S(A)$ denotes the average richness of domains with area $A$, \citet{arrh21:95} found that the power law $S=cA^z$ held for plants on a set of islands, while \citet{glea22:158} found that the semi-log law $S=a+b\log(A)$ held for plants in patches of a single vegetation type. Both of these laws have the form
\begin{equation}\label{SARgen}
S = f(A),
\end{equation}
where the function $f$ is explicitly given in a short formula with few parameters.

Since then many other likewise simple forms have been found to fit data on $S$ versus $A$. The review by \cite{tjor03:827} lists 14~such functions in its two main categories and mentions several more. The most commonly used remain the power law and the semi-log law: for example, \citet{draketal06:215} found 796~and 506~cases, respectively. Such a wealth of results demand interpretation, for which one requires a theoretical framework. Usually, this is along the lines elegantly presented by \cite{may75:81}. The starting point is to assume known the species abundance distribution (SAD) and the total area (i.e.\ the largest area over which the SAR is to be applied). Next, it is assumed that the density of individuals is uniform throughout the area and that they are independently distributed. The expected number of species in a given area $A$ then follows from a probabilistic calculation. In other words, this approach explains SARs from SADs. Normally the latter are derived from ecological processes using a second theoretical framework. One is thereby enabled to interpret differences in the parameters or even in the form of equation~\ref{SARgen} in terms of underlying ecological processes. A high point of this approach is Rosenzweig's magisterial synthesis~(\citeyear{rose1995}), which uses power law SARs interpreted in terms of island biogeography. Other theories for which this approach works include metapopulation dynamics~\citep{hansgyll97:397,ovashans03:903} and Hubbell's neutral theory~\citep{volketal03:1035,bell00:606,chav04:241}.

A completely different modelling paradigm is due to Harte and colleagues~\citep{hartkinz97:417,hartetal99:334}. They invert May's sequence, and derive an SAD from an SAR. In their models, the spatial domain is a rectangle with area $A_0$, and it is repeatedly refined by a sequence of bisections into increasingly smaller rectangles. The SAR is applied at each bisection, thus specifying the number of species retained in each half. In their papers, Harte and colleagues use the power law $S=cA^z$, and this implies that the proportion of species retained in each half is $a=2^{-z}$. By further assuming that the bisection cascade ends with $2^{m}$ rectangles of size $A_0/2^m$ and that each such rectangle contains one individual, an SAD can be derived~\citep{hartetal99:334,martgold02:032901}.

It should be noted that the SAD derived from $S=cA^z$ in the Hartean scheme is \emph{not} the SAD from which May derives $S=cA^z$. The reason appears to be that in May's derivation each individual is indpendently placed, so that spatial autocorrelation is zero by assumption, while in Hartean models individual positions within species are correlated, in that individuals of a rare species are very likely to occur near each other, and hence spatial autocorrelation within species cannot be zero.

The Hartean approach is attractive for several reasons. Firstly it directly gives species richness as a density which varies according to scale. Put differently, it provides a simple way to simulate species richness across many scales. Secondly, it is invariant with respect to scale. That is, the relationship between two patches depends only on their relative areas, not on their absolute areas. Thirdly, it is extremely simple. For example, it uses only one parameter, whereas May uses two (actually three, but any one parameter can be derived from the other two). As a purely descriptive tool it has much to recommend it. However, its very simplicity counts against it. It has been pointed out that the bisection cascade creates a very limited set of areas, leading to the claim that the model does not apply to areas not included in this small set~\citep[on which more below]{madd04:616}. Also, and perhaps more importantly, it is explicitly a model of average richness. This limitation is of course shared by all models in the form of equation~\ref{SARgen}. Nevertheless, variability in richness is an obvious fact and indeed one of the main objects of study in biodiversity~\citep{connmcco01:397} and is known to be scale-dependent~\citep{rahb05:224,paut07:16}. It also varies in time~\citep{gonz00:441,adlelaue03:749}. A description of species richness with appropriate scaling properties for the average richness as well as the variation in richness is likely to be very useful.

In this paper we present such a model: MFp1p2. It is a simple generalisation of the Hartean approach, where instead of assuming that both halves of a bisection retain the proportion $a$, we assume instead that they differ, with the richer half retaining the proportion $p_1$ and the poorer half retaining $p_2$ of the species in the bisected rectangle. As a result, the rectangles with given area shows a range of richnesses. In other words, MFp1p2 gives the SAR as a set-valued function: $S$ at a given $A$ takes a set of values. We give an exact expression for the frequency distribution of these values, and derive a number of consequences: self-similarity, a simple graphical characterisation, numerous power-law scalings, and the prediction that richness and endemism are positively correlated.

What about objections to the original Hartean model \citep[e.g.][]{madd04:616}? Do they not also hold in this case? Yes, but we do not believe they invalidate the use of either MFp1p2 or its Hartean parent. Firstly, these models are descriptive, not causal. Their value lies in their ability to summarise complex reality in a few parameters. They need merely do that with sufficient accuracy. In this respect, they resemble null models and neutral models. In spatial ecology at present such descriptions are indispensable, even when not very accurate. Without them, we have only the fully detailed maps of distributions and richness. Secondly, the limitations of the constructions used to explain these models can in many ways be mitigated by interpretative strategies and extensions. We explore both of these points more fully in the discussion.

\section*{MF\lowercase{p1p2}: A TWO-PARAMETER MULTIFRACTAL MODEL}

\begin{figure} 
\includegraphics[width=10cm]{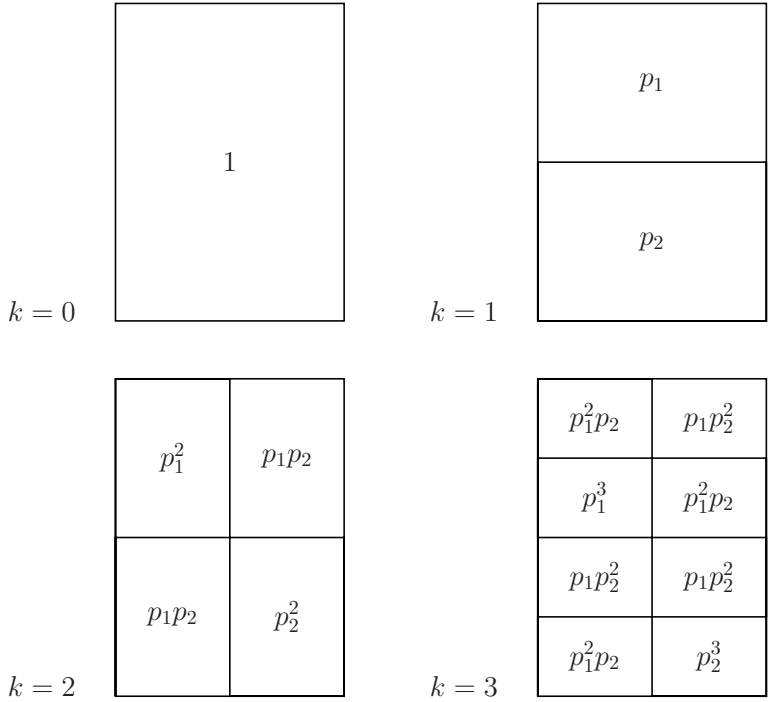}
\caption{Constructing MFp1p2: stages $k=0$, 1, 2 and~3 of the multiplicative cascade. At each stage, there are $2^k$ bisections. $p_1$ is the fraction of richness inherited by the richer half, $p_2$ is the fraction of richness inherited by the poorer half. Position of richer half is randomly chosen among the two possibilities at each bisection.}
\label{makingp1p2}
\end{figure}

This exposition refers to Figure~\ref{makingp1p2}. Consider a rectangle with area $A_0$ that contains $S_0$ species. Bisect the rectangle into two congruent rectangles of area $A_1=A_0/2$, then bisect each of these with a line at right angles to the previous bisector to reach four rectangles of area $A_2=A_0/4$. Continue this process, reaching $2^k$ rectangles of area $A_k=A_0/2^k$ after $k$ bisections.

\begin{figure} 
\includegraphics[width=10cm]{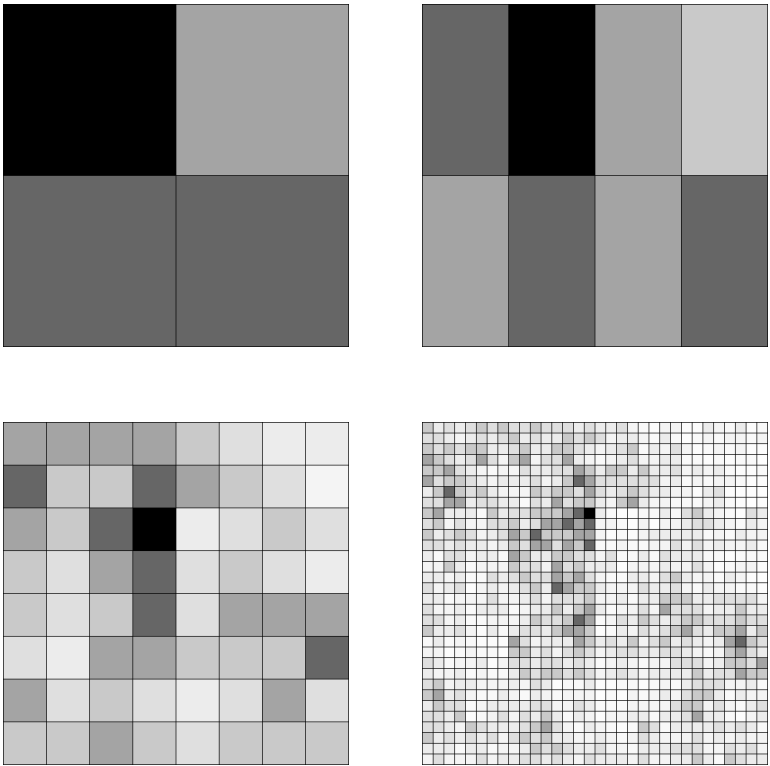}
\caption{Realisations of MFp1p2 with $p_1=1$, $p_2=0.6$, for $k=2$, $k=3$, $k=6$ and $k=10$. Density scale is the same in all four cases: the maximum richness is $p_1^2=1$, which appears as black areas (squares/rectangles) in each of the four examples, and the minimum richness is $p_2^{10}\approx 0.006$, which appears only in the $k=10$ example.}
\label{p1p2reals}
\end{figure}

As each parent rectangle is split in two, spatial heterogeneity in species richness implies that its offspring inherit unequal amounts of species. Denote by $p_1$ the fraction of species inherited by the richer half and by $p_2$ the fraction  inherited by the poorer half. For example, after the bisection of $A_0$ the richer half contains $p_1S_0$ species, and the poorer half contains $p_2S_0$ species. Since every species in the larger rectangle must occur in at least one of the smaller rectangles, we have $p_1 + p_2\geq1$. At each stage, the position of the richer half is randomly assigned. Typical realisations are shown in Figure~\ref{p1p2reals}.

Notice that $p_1$ and $p_2$ remain constant throughout the construction. After $k$ bisections the heterogeneity is captured in the binomial $(p_1+p_2)^k=\sum_{j=0}^k\binom{k}{j}p_1^{k-j}p_2^j$. Species richness takes the $(k+1)$ values $p_1^{k-j}p_2^jS_0$ for $j=0$~to~$k$, and each level of richness occurs in $\binom{k}{j}$ of the $2^k$ rectangles with area $2^{-k}A_0$. If one denotes by $P_k(S=n)$ the proportion of rectangles with area $2^{-k}A_0$ that contain $n$ species, then we can restate this result as the species richness distribution
\begin{equation}\label{SRD}P_k(S=S_0p_1^{k-j}p_2^j) = \binom{k}{j}/2^k. \end{equation}

Of course, Equation~\ref{SRD} has no spatial reference, and therefore is not a complete representation of MFp1p2; it gives only the frequencies of the richness at each level. The full MFp1p2 is better thought of as a density where extremes of richness are rare, the richness distribution is unimodal at every scale, and near any cell with high richness there will be cells with low richness, but cells with high richness will tend to be near other cells with high richness. 

MFp1p2 is not very different from the 2-dimensional multifractal constructed by \citet{stanmeak88:405}; a much-cited example of the increasing use of multifractals in many fields including porous media~\cite{perretal06:284}. At every second step in constructing MFp1p2 one reaches a step in their construction. What is new here is firstly that $p_1$ and $p_2$ have a natural interpretation in terms of species richness and secondly the explicit distribution of equation~\ref{SRD}.

A classical  way to characterize a multifractal measure is based on the calculation of the Renyi dimensions of a multifractal object. ~\citet{hart01} gives the definition as 

\newcommand{\mystrut}{\rule[-2ex]{0em}{8ex}}

\begin{equation}\label{rendef}
D(q) = \begin{cases} \displaystyle \lim_{r\rightarrow0}\frac{1}{q-1}\frac{\log_2\sum_{i=1}^{N(r)}p_i^q(r)}{\log_2 r} \qquad\text{ if $q\neq1$} \\ \mystrut\displaystyle
\lim_{r\rightarrow0}\frac{\sum_{i=1}^{N(r)}p_i(r)\log_2{p_i(r)}}{\log_2 r} \qquad\text{ if $q=1$},\end{cases}
\end{equation}

\noindent where $r$ is a characteristic length, it takes $N$ identical boxes with length $r$ to cover the object, and $p_i$ is the proportion of the density in box~$i$. 

Substituting the probabilities from Equation~\ref{SRD} into Equation~\ref{rendef} one obtains the following expression for the Renyi dimensions of MFp1p2:
\begin{equation}\label{rendim}
D(q)  = \begin{cases}
 \displaystyle \frac{2}{1-q}\log_2\frac{p_1^q+p_2^q}{(p_1+p_2)^{q}} \qquad \text{ if $q\neq1$} \\
\\ 
  \displaystyle \log_2(p_1+p_2)^2- \frac{p_1\log_2(p_1)^2-p_2\log_2(p_2)^2}{p_1+p_2} \\
\hfill \text{ if $q\neq1$}.
        \end{cases}
\end{equation}
This in fact holds for finite $k$, rather than merely in the limit of large $k$ (see Appendix).

\subsection*{Reparameterisation using $a$ and $b$}

We now turn to an alternative and equivalent way to see MFp1p2, via $a=(p_1+p_2)/2$ and $b=p_1/p_2$.

\begin{figure}
\includegraphics[width=10cm]{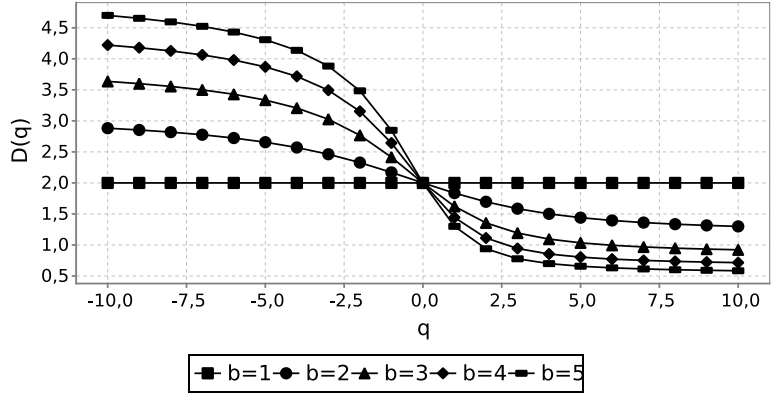}
\caption{Typical $D(q)$ for various values of the variability scaling $b$. The flat line when $b=1$ is the spatially invariant case of \citet{hartetal99:334} where $p_1=p_2$; as $b$ grows the spatial variability becomes more pronounced and the curves depart more and more from a straight flat line.}
\label{Dqcurves}
\end{figure}

Firstly, MFp1p2 is a generalisation of the density constructed in \citet{hartetal99:334}, as follows. When $a=p_1=p_2$, our construction reduces to theirs. When $p_1\neq p_2>0$, we use the mean richness ratio $a = (p1+p2)/2$. Define mean richness $\bar{S}_k$  after $k$ bisections as the sum of all the richnesses divided by $2^k$. Clearly, the mean richness satisfies the recurrence $\bar{S}_{k+1} = (p_1+p_2)\bar{S}_k/2 =  a\bar{S}_k$. This recurrence relation immediately implies the power law $\bar{S}=cA^z$, where $z=\log_2{1/a}$ and $c=S_0/A_0^z$, exactly in the same way as in \citet{hartetal99:334}. Hence the density of species richness they construct purely as a mean field model is exactly the trend of richness in MFp1p2. 

Secondly, the variability in $S$ at a given $a$ depends entirely on $b=p_1/p_2$: after $k$ bisections, the ratio between the richest and poorest rectangles of size $A_0/2^k$ is $b^k$. The variability increases by the factor $b$ at every step, so $b$ governs the scaling in variability. Interestingly, $D(q)$ depends only on $b$:
\begin{equation*}\label{Dqoneparam}
D(q)  = \begin{cases} \displaystyle\frac{\log_2(b^{2q}+1)-\log_2(b+1)^{2q}}{1-q} & \text{ if $q\neq1$} \\ \displaystyle
  \log_2(b+1)^2-\frac{b\log_2(b)^2}{b+1} & \text{ if $q=1$}.\end{cases}
\end{equation*}

In sum, $a$ gives the scaling of the trend and $b$ gives the scaling in the variability. Typical $D(q)$ curves for MFp1p2 are shown in Figure~\ref{Dqcurves}

\subsection*{Some properties of MFp1p2}
\begin{enumerate}

\item Many statistics of MFp1p2 have power laws. Above we derived $\bar{S}=cA^z$, the usual interpretation of the species-area relationship. Similar expressions can be derived for the variance, minimum, mode and maximum of $S(A)$.

\item MFp1p2 is exactly self-similar in the following sense. As $k\longrightarrow\infty$, every rectangle at level $k$ is a scaled copy of the initial rectangle, provided we consider richness relative to overall richness and ignore the placement of the two halves resulting from each bisection.

\item MFp1p2 has a characteristic triangular footprint on a log-log scatterplot, as in shown in Figure~\ref{footprint}. 

\begin{figure}

\includegraphics[width=10cm,angle=0]{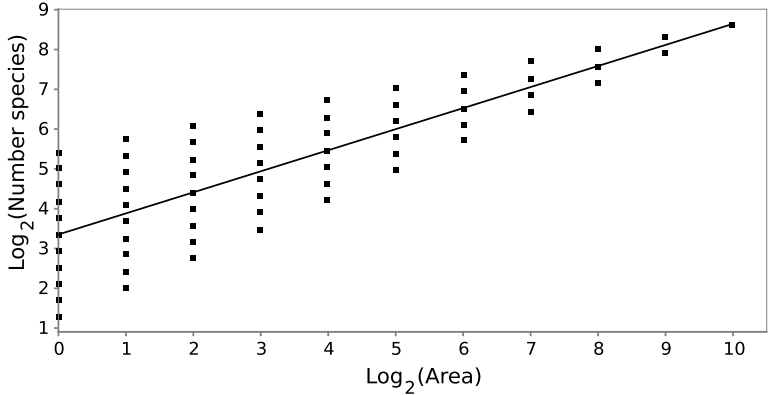}

\caption{
The triangular footprint of MFp1p2 on a log-log scatterplot. The non-vertical edges have slopes~$\log_2(p_1)$ and~$\log_2(p_2)$, the horizontal spacing is~1~($=\log_2(2)$) and the vertical spacing is~$\log_2(b)$. The straight line is the classic Arrhenius power law with $a=(p_1+p_2)/2$ and goes through the mean $\bar{S}$ at every $A$.
\label{footprint}}
\end{figure}

This illustrates graphically the way MFp1p2 generalises the classical species-area power law: the triangular footprint collapses to the classic straight line when $b=1$; the mean $\bar{S}$ of MFp1p2 gives the classic straight line.

In the example below, we show that it can match real data.

\item MFp1p2 predicts a positive correlation between endemism and richness. Let us suppose that $A_0$ contains $E_0$ endemics (species that do not occur outside $A_0$) and that endemics by themselves also divide according to $p_1$ and $p_2$. Then the proportion of endemics that occur in the richer half with area $A_0/2$ is $1-p_2$, because those species do not occur in the poorer half. That is, these species are endemics not only of the whole area $A_0$ but also of the richer half. We define $e_1 = 1-p_2$ (respectively $e_2=1-p_1$) as the proportion of endemics that are endemic to the richer (respectively poorer) half of a a bisection, and denote by $P_k(E=n)$ the probability that a rectangle with area $A_0/2^k$ contains $n$ endemic species.  We then obtain the endemics distribution $P_k(E=E_0e_1^{k-j}e_2^j) = \binom{k}{j}/2^k$. Because $e_1>e_2$ when $p_1>p_2$, the richest rectangles at a given $k$ are precisely those with the largest number of endemic species.

Just as for richness, statistics for endemism have power laws. For example, for mean endemism the exponent is $z_E=\log_2(1/a_E)$, where $a_E=1-a$; here again we recover a result first given by by Harte and co-workers~\cite{hartkinz97:417}.

\end{enumerate}

\subsection*{Illustration: Proteaceae richness in the Cape Floristic Region}

The Cape Floristic Region (CFR) of the south-western corner of Africa is the smallest of the planet's six floristic regions recognized in \citet{takh86}, among which it has the highest endemism. Proteaceae are the signature plants of fynbos, a vegetation type almost entirely confined to the CFR. The CFR is made up of several biomes, one of which is dominated by fynbos and hence is named the fynbos biome. Data on species richness of Proteaceae was kindly provided by Tony Rebelo of the Protea Atlas Project~\cite{rebe91}. These data are derived from approximately 60~000 site record sheets. These were filled in as follows. Firstly, the centre of the site was recorded using the GPS, then a roughly circular area with this centre and roughly 500 metres in diameter was designated, and finally all Proteaceae species found there were recorded on a standard form (as well as some other data irrelevant here; some pre-existing data were also used).

\begin{figure}
\includegraphics[width=7.2cm,clip]{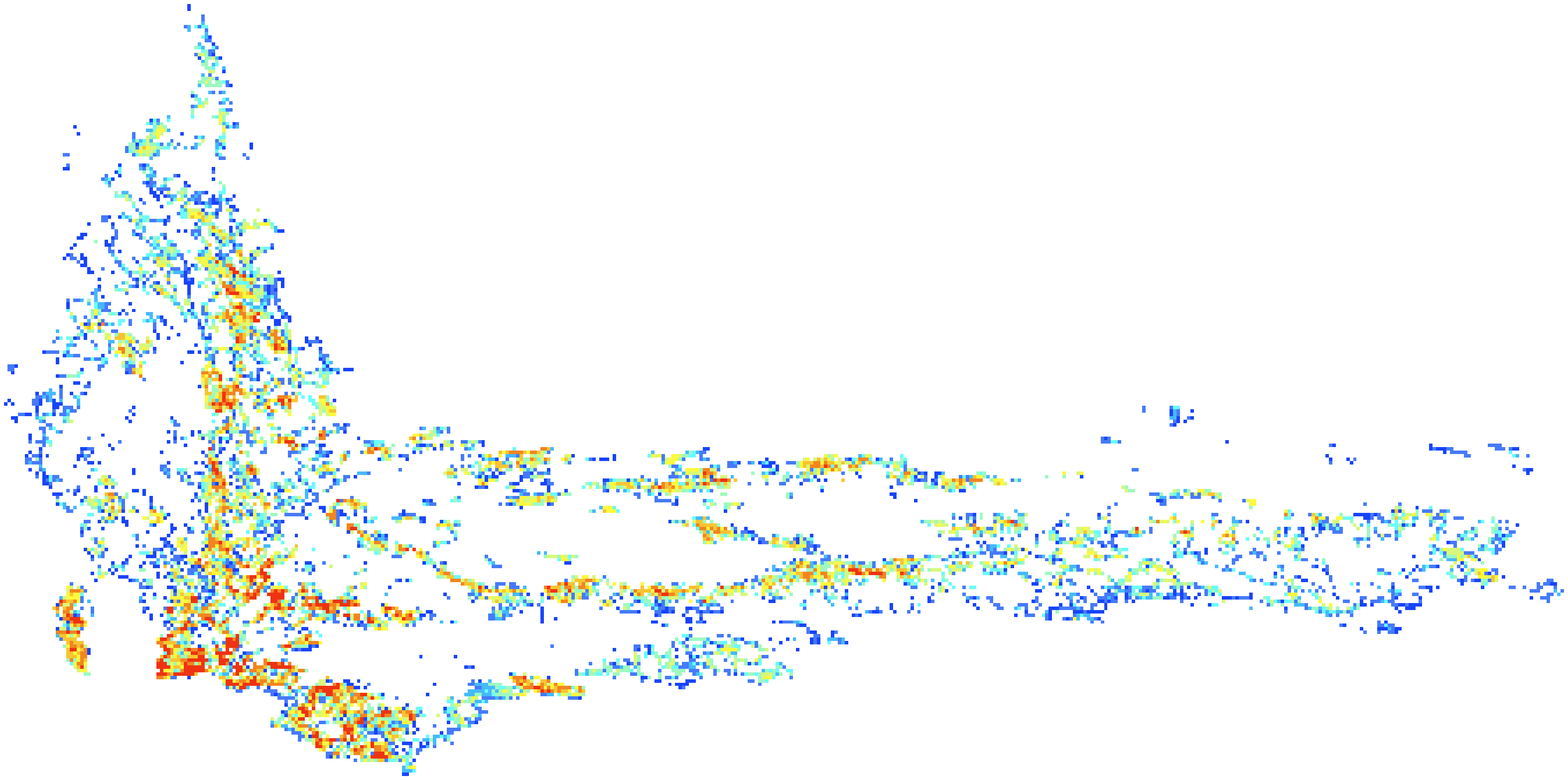}
\includegraphics[width=7.2cm,clip]{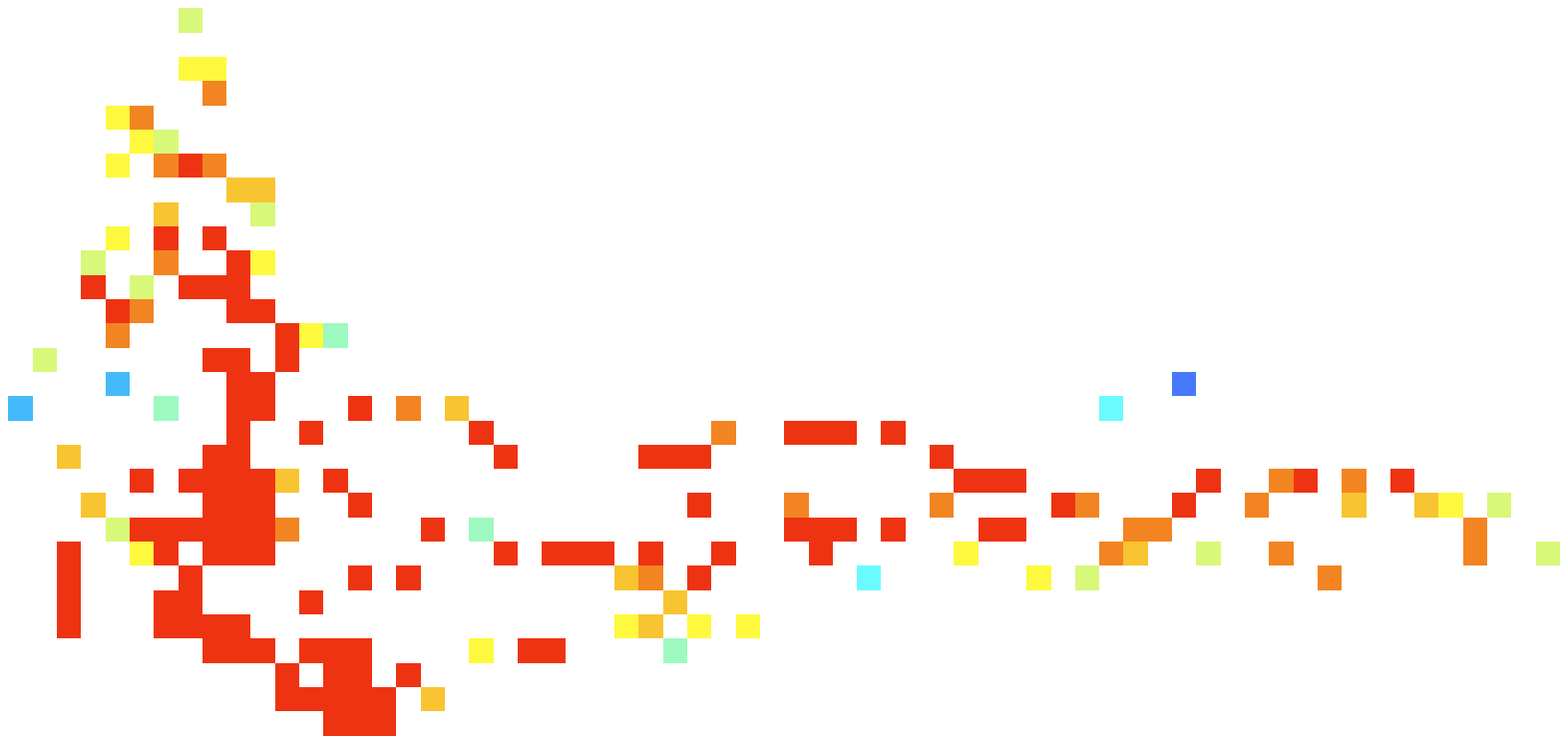}
\caption{Maps of richness of Proteaceae species in Cape fynbos. The first shows the finest available scale in the data: each cell is $1'\times1'$. In the second, each cell is $7'\times7'$. Cells are coloured according to number of species they contain, with the same colourmap from poorest (blue) to richest (red) in both maps; non-fynbos cells are blank.}
\label{PAmaps}
\end{figure}

In this paper, we consider a rasterization of the CFR by lines of longitude and latitude $1'$ apart. The resulting cells are very nearly rectangular, approximately 1~km by 1.5~km. The fynbos biome is here defined by protea presence: for the purposes of this illustration it consists of those cells that have at least one occurence of a Proteacaeae species recorded in the Protea Atlas. We call them fynbos cells, of which there are approximately 9~000 in the atlas. We wish to study the SAR of Proteaceae in fynbos. In other words, we need data giving area $A$ and richness $S$ samples from the fynbos biome. The samples were constructed as follows. Each fynbos cell was taken as the centre of a sequence of nested cells of size $d$~minutes by $d$~minutes. The smallest cell is of course $1'\times1'$, the next size is $3'\times3'$ and so on. The full set of $d$ are 1, 3, 7, 15, 31, 63, 127, 255, 511, 1023. The fynbos area $A$ inside a given cell is then simply the number of $1'\times1'$ fynbos cells it contains, and the count $S$ is from the full list of species of these combined cells. Thus each $1'\times1'$ fynbos cell is at the centre of a nested sequence of fynbos areas, each with its own $(A, S)$ values. At the largest scale, each cell contained all 9~000 fynbos cells. See Figure~\ref{PAmaps} for maps of richness using cells at two different scales.

\begin{figure}
\includegraphics[width=7.2cm,clip]{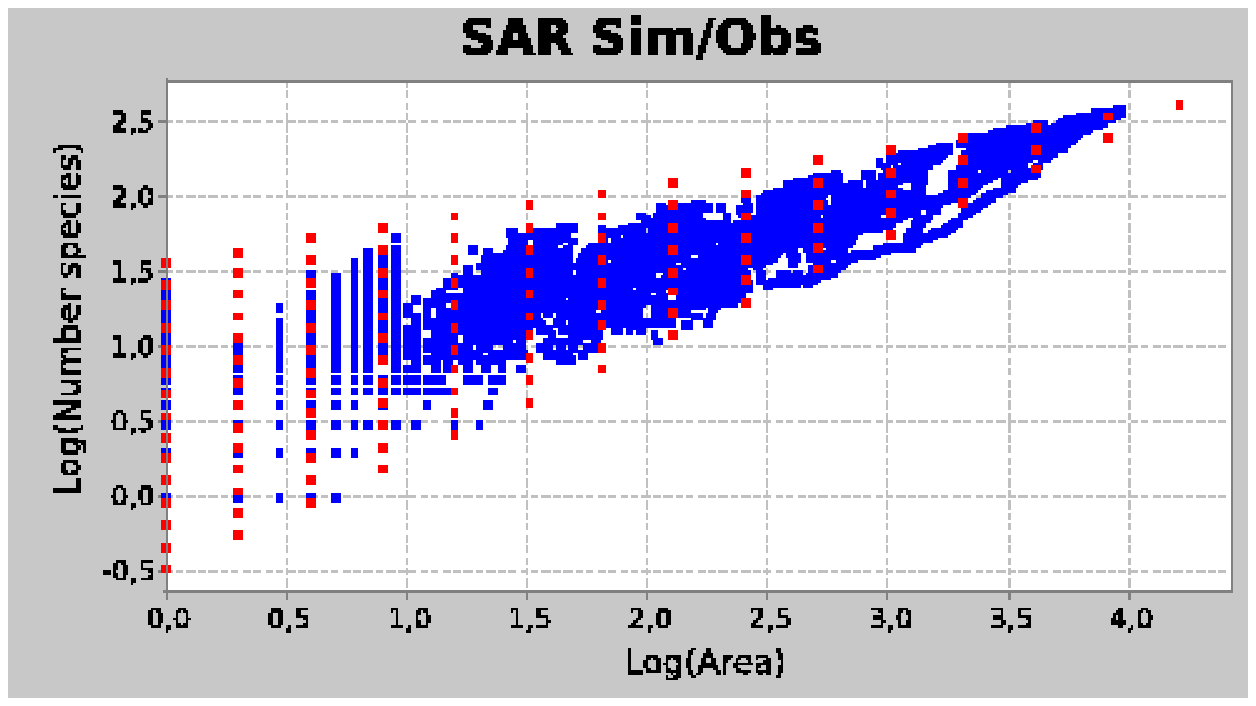}
\includegraphics[width=7.2cm,clip]{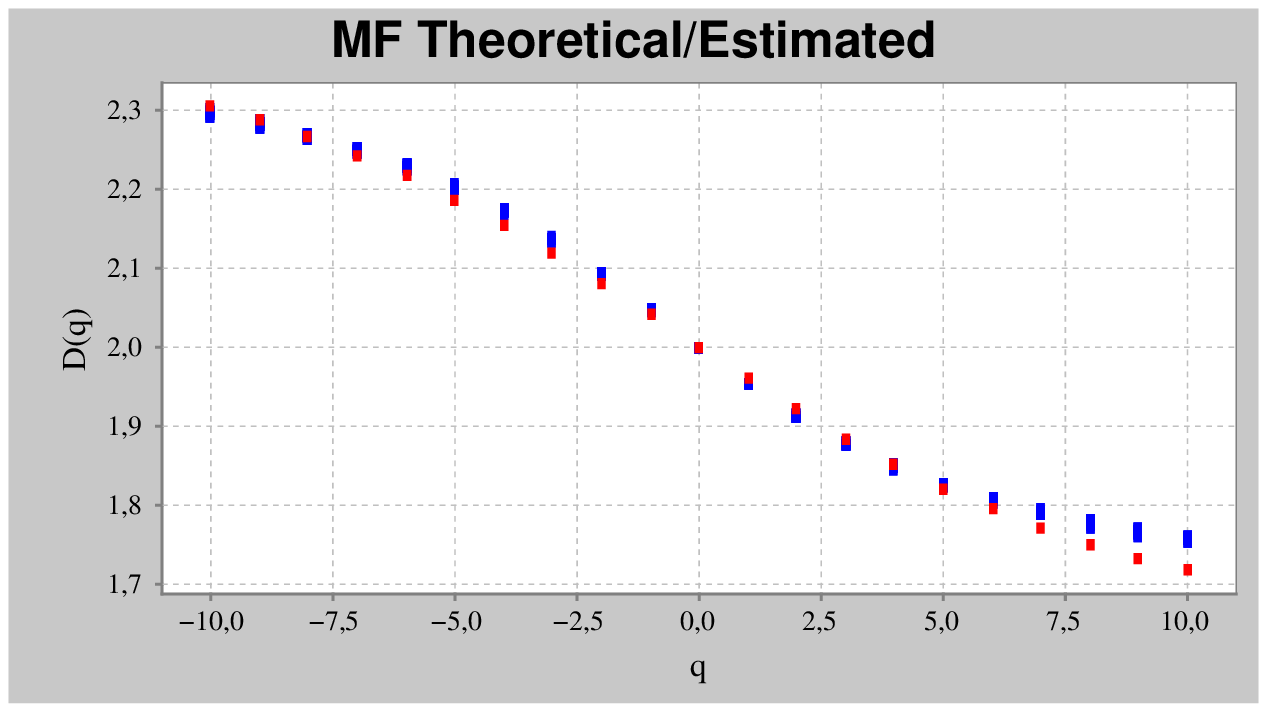}
\caption{Example of application of MFp1p2: scaling of mean and variability of Proteaceae richness in Cape fynbos. (a)\ Triangular footprint of species richness, with an MFp1p2 triangular footprint superimposed. Area is in number $1'\times1'$ cells; richness is number of species. Axes are scaled logarithmically. The blue dots are $(A, S)$ pairs from the Protea Atlas, while the red dots are the triangular footprint of MFp1p2 (with $p_1=0.83$, $p_2=0.61$; equivalently $a=0.72$, $b=1.36$). (b) Spectra of Renyi dimensions: species richness of Cape Proteaceae (blue dots) and an MFp1p2 spectrum (red dots, $p_1=0.83$, $p_2=0.61$; equivalently $b=1.36$).}
\label{ProtSAR}
\end{figure}

Figure~\ref{ProtSAR} shows that the $(A,S)$ data described above have a roughly triangular log-log footprint, and some values of $p_1, p_2$ that allow one to superimpose an MFp1p2 footprint that fits it quite well. Of course, MFp1p2 has only a small number of discrete values for $(A,S)$ pairs, whereas the real data appear quite densely packed. Thus MFp1p2 matches the pattern of the triangular footprint, but not all its details. It succeeds quite well in characterising the trends of both the mean richness and the variation in richness over many orders of magnitude of area. Using an algorithm for estimating Renyi dimensions from data (Perrier, E. and Laurie, H., unpublished work), the figure also shows that the data have a spectrum of Renyi dimensions very similar to that of an MFp1p2 model.

\section*{DISCUSSION}

The variability of species richness at given area has been neglected in studies of SARs. We believe the reasons for this no longer hold. It is obviously very onerous to collect and process sufficient data to establish not only the trend but also the variability in $S$ at a given $A$. However, over time large data sets have been accumulated and with the developments in computing they can be handled. As to theory, the construction of species density via an algorithm~\citep[pioneered by][]{hartetal99:334} avoids two limitations of the derivation of SARs from SADs by probabilistic arguments. Firstly, no assumption is needed about spatial autocorrelation. As pointed out above, non-zero spatial autocorrelation emerges automatically, and if need be additional assumptions are easily added~\citep[as in][]{martgold06:10310}. Secondly,  algorithmic simulation allows direct modelling of processes at particular scales. By contrast, the allocation problem for the scale of spatial variation of smoothly varying densities is unsolvable~\citep{cres93}: that is, from data alone one cannot identify the exact contribution of variation at each scale to the overall variation if the density varies smoothly.

Let us note three recent papers that develop Hartean theory of spatial structure in SARs in ways quite different from MFp1p2. All three embrace the SAD. \citet{hartetal05:179} introduces the concept of spatial abundance distribution, by which the overall SAD is distributed across the cells created in a bisection cascade. Here, besides the parameter $a$ of~\citet{hartetal99:334}, the abundances $n_0$ of each species are used. However, the traditional derivations do not apply, because of the spatial autocorrelation necessarily introduced by the bisection cascade and the parameter $a$. Instead, the analysis follows from a  further assumption called HEAP, which we do not treat here. Something similar is done by~\citet{martgold06:10310}: they also start with the SAD as given and allocate the individuals via a bisection cascade. However, instead of the parameter $a$ and HEAP, \citet{martgold06:10310} assume that individuals in each species are clustered. The generality of this assumption precludes analytical results, but their computer experiments are extensive and show convincingly that realistic SARs do result from a variety of clustering assumptions. \citet{conletal07:169} is rather more preliminary. Here, the SAD is again the starting point for an allocation of species to a cascade of smaller and smaller regions. However, the geometry is generalised from bisection to $n$-section, whereby a given rectangle is divided into $n>=2$ subregions at each step. The individuals in a parent region are distributed one by one to the subregions at each dissection step by a probabilistic rule. The authors show that this leads to  coherent and reliable simulations, because for a given setting of the parameters, the final state of the system is independent of the exact sequence by which it was reached. They also study some statistical issues in parameter estimation with reference to three data sets.

These papers  differ from MFp1p2 in three ways. Firstly, they treat spatial variation indirectly. Secondly, analysis of spatial variation is not pursued. Thirdly, they need at least as many parameters as there are species. By contrast, spatial variation is directly modelled and quantified in MFp1p2 using only two parameters. Thre usual trade-offs apply: loosely speaking, models with more parameters fit the data more closely, with less confidence in estimates of parameter values and more sensitive predictions.

A particular strength of MFp1p2 is that the parameter $b=p1/p2$ directly quantifies spatial variability across scales. To our knowledge, it is the first SAR model that quantifies variability independently of mean richness. We suggest that $b$ may be a very useful summary statistic.

The most obvious prediction of MFp1p2, as clearly demonstrated by the log-log footprint, is that variability in $S$ increases as $A$ shrinks. This is in marked contrast with the one published result on spatial variability the we are aware of, namely the random placement model~\citep{cole81:191}, where $\text{Var}(S)$ goes to zero with $A$. The Proteaceae data shown above indicate that in at least some cases the prediction of MFp1p2 is correct.

The ability directly to evaluate $b$ as well as $a$ from data should allow comparisons of variability in $S$ along the same lines previously used for the mean of $S$~\citep{rose1995}. It would be very interesting if situations are found where the values of $a$ are similar while the values of $b$ differ, or vice versa. It would be even more interesting if $a$ and $b$ were strongly correlated. 

We turn now to the objection \citep{madd04:616} that bisection cascades are fatally impoverished: they furnish far too few values of $A$ and far too few instances of domains with a given $A$. It is true that only $k+1$ values for $A$ occur from $A_0$ to $A_k$, and also true that only a few of the possible rectangles that can be formed from the recangles of size $A_k$ that actually occur in the cascade. But the point is to describe the SAR, not to recreate it in detail by generating its value for all possible geometries at every $A$ in a continuous range. That is, the aim is to summarise the behaviour of $S$ versus $A$. In the case of Cape Proteaceae (see above), the log-log footprints of data and model quite clearly agree as to the trend of both the mean and of the variability. Here, MFp1p2 appears to do a good descriptive job. The spectra of Renyi dimensions also match acceptably.

The Cape Proteaceae example also shows that real data have quirks that a simple model could not possibly catch. Generalisations and extensions are necessary. We note that two possibilities remain quite simple. The first is that, instead of being  constant, $p_1$ and/or $p_2$ could be functions of $A$, thus introducing variation across scales. This might involve only a few extra parameters. The SAR would no longer be a power law, but the resulting density would still be multifractal (albeit not self-similar even in the weak sense used above). The second is that the shape of $A_0$ and indeed of all the $A_i$ need not be rectangular: any shape, no matter how dissected or disconnected, can be bisected in many ways. Thus bisection cascades can be constructed for the typically complicated domains on which ecology is studied, and in many ways.

Some immediate research tasks concerning MFp1p2 are obvious. Application should be tested by estimating $a$ and $b$ for more sets of real data and studying the reliability of such estimates. We already have indications for several data sets that reliable estimates are possible (this is work in progress). The results of bisection cascades using various bisectors and rasterisations should be tested for consistency. There are also interesting open theoretical questions. One is the lack of a rigorous basis for inferences concerning SARs, because the correct probability distribution of domains with area $A$ is unknown. The distribution cannot be based for all sets with area $A$, as the majority of such sets are ecologically meaningless. On the other hand, it cannot be based on just the convex sets, nor on the connected sets. Without a set of $A$ to sample from, one cannot rigorously infer reliability, or indeed make any other inference. Note that this difficulty parallels the problem of rigorous estimation of the actual (as opposed to observed) number of species, which is still the subject of research~\citep{haasetal06:135}.

\bibliography{LaurPerr}

\section*{APPENDIX: Derivation of formula for $D(q)$}

First consider the case $q\neq1$. By definition, $D(q) = \frac{\log_2\sum_i^{N(r)}p_i^q(r)}{(1-q)\log_2 r}$. We need the $p_i$, which are proportions of the total density. From equation~\ref{SRD} we know there are $\binom{k}{j}$ rectangles with richness $p_1^{k-j}p_2^jS_0$ after $k$ bisections. The sum of all richness is $(p_1+p_2)^kS_0$. In other words, $$\displaystyle\sum_{i=1}^{N(r)}p_i^q(r) = \sum_{j=0}^k\binom{k}{j}\left(\frac{p_1^{k-j}p_2^j}{(p_1+p_2)^k}\right)^q = \frac{(p_1^q+p_2^q)^k}{(p_1+p_2)^{kq}}$$.

Each rectangle has an area proportional to $2^{-k}$, so one may use $r=\sqrt{2^{-k}}$ as the characteristic length. It follows that $k\rightarrow\infty$ as $r\rightarrow0$. Hence we have $$\displaystyle D(q)=\lim_{k\rightarrow\infty}\frac{1}{q-1}\frac{\log_2\bigl(\frac{p_1^q+p_2^q}{(p_1+p_2)^q}\bigr)^k}{\log_{2}2^{-k/2}} = \lim_{k\rightarrow\infty}\frac{1}{q-1}\frac{k\log_2\frac{p_1^q+p_2^q}{(p_1+p_2)^q}}{-k/2}$$.

We see that $k$ cancels, so there is no need to take the limit for large $k$, and the first expression in equation~\ref{rendim} drops out. The second expression follows from the first by L'Hospital's rule.

\end{document}